\documentclass[useAMS]{mn2e}
\usepackage{times}
\usepackage{txfonts}
\usepackage{lscape}

\usepackage{graphicx}
\usepackage[dvips]{color}

\newcommand{\reference}{\bibitem}

\def\aap{A\&A}
\def\apj{ApJ}

\newcommand\plotone[1]{\centering\includegraphics[width=0.65\hsize]{#1}}

\def\beq{\begin{equation}}
\def\eeq{\end{equation}}
\def\bey{\begin{eqnarray}}
\def\eey{\end{eqnarray}}

\def\tEfourp{t_{\rm E,4p}}
\def\tEfivep{t_{\rm E,5p}}
\def\tE{t_{\rm E}}
\def\tEout{t_{\rm E, out}}
\def\tEin{t_{\rm E, in}}
\def\fSin{f_{\rm S, in}}
\def\fSfivep{f_{\rm S, 5p}}
\def\fS{f_{\rm S}}
\def\moa{MOA-2003-BLG-37}
\def\Iout{I_{\rm out}}
\def\Iin{I_{\rm in}}
\def\IS{I_{\rm S}}
\def\tauout{\tau_{\rm out}}
\def\tauin{\tau_{\rm in}}

\def\u0{u_0}

\def\mI0{m_{I,0}}

\input epsf
\begin{document}

\title[Blending in Gravitational Microlensing Experiments]{Blending in
  Gravitational Microlensing Experiments: Source Confusion And Related
  Systematics}

\author[Smith et al.]
{Martin C. Smith$^{1}$, Przemys{\l}aw Wo\'zniak$^2$, Shude Mao$^3$,
  Takahiro Sumi$^{4,5}$
\thanks{e-mail: msmith@astro.rug.nl; wozniak@nis.lanl.gov;
  smao@jb.man.ac.uk; sumi@stelab.nagoya-u.ac.jp}
\\
\smallskip
$^1$Kapteyn Astronomical Institute, University of Groningen, P.O. Box 800, 9700 AV Groningen, Netherlands \\
$^2$Los Alamos National Laboratory, MS-D436, Los Alamos, NM 87545, USA \\
$^3$Univ. of Manchester, Jodrell Bank Observatory, Macclesfield,
Cheshire SK11 9DL, UK \\ 
$^4$Princeton University Observatory, Princeton, NJ 08544-1001, USA\\
$^5$Solar Terrestrial Environment Laboratory, Nagoya University,
Nagoya, Aichi 464-8601, Japan\\
}
\date{Accepted ........
      Received .......;
      in original form ......}

\pubyear{2007}

\maketitle
\begin{abstract}
Gravitational microlensing surveys target very dense stellar
fields in the local group. As a consequence the microlensed source stars
are often blended with nearby unresolved stars. The presence of `blending'
is a cause of major uncertainty when determining the lensing properties of
events towards the Galactic centre.  
After demonstrating empirical cases of blending we utilize Monte Carlo
simulations to probe the effects of blending.
We generate artificial microlensing events using an $HST$ luminosity
function convolved to typical ground-based seeing, adopting a range of
values for the stellar density and seeing. Microlensing light curves
are generated using typical sampling and errors from the second phase
of the Optical Gravitational Lensing Experiment. We find that a
significant fraction of bright events are blended, contrary to the
oft-quoted assumption that bright events should be free from blending.
We probe the effect that this erroneous assumption has on both the
observed event timescale distribution and the optical depth, using
realistic detection criteria relevent to the different surveys.
Importantly,
under this assumption the latter quantity appears to be reasonably
unaffected across our adopted values for seeing and density. The
timescale distribution is however biased towards smaller values, even
for the least dense fields.
The dominant source of blending is from lensing of faint source stars,
rather than lensing of bright source stars blended with nearby fainter
stars.
We also explore other issues, such as the centroid motion of blended
events and the phenomena of `negative' blending.
Furthermore, we breifly note that blending can affect the
determination of the centre of the red clump giant region from an
observed luminosity function. This has implications for a variety of
studies, for example mapping extinction towards the bulge and attempts
to constrain the parameters of the Galactic bar through red clump
giant number counts.
We conclude that
blending will be of crucial importance for future microlensing
experiments if they wish to determine the optical depth to within 10
per cent or better.
\end{abstract}


\section{Introduction}
\label{sec:intro}

Gravitational microlensing is maturing into an important astrophysical
technique with diverse applications to Galactic astronomy, such as
probing the dark matter content of the inner Galaxy 
(see, e.g., the following review articles: Paczy\'nski 1996; Mao 1999;
Evans 2003).
Thousands of microlensing events have been discovered. The vast majority
of these are towards the Galactic centre and many were identified in
real-time, for example by the
OGLE\footnote{http://www.astrouw.edu.pl/\~{}ogle/ogle3/ews/ews.html}
(Udalski 2004) or 
MOA\footnote{http://www.massey.ac.nz/\~{}iabond/alert/alert.html}
(Bond et al. 2001) alert systems.
The microlensing probability (known as the optical depth, $\tau$)
towards the Galactic centre probes the mass distribution along the line
of sight. The earliest determinations yield optical depths
(Udalski et al. 1994a; Alcock et al. 1997a; Alcock et al. 2000) that
are significantly higher than the theoretical predictions (e.g., Zhao \&
Mao 1996; Binney, Bissantz \& Gerhard 2000; Evans \& Belokurov 2002;
Han \& Gould 2003). 
More recent determinations yield both lower values (Popowski et
al. 2005; Hamadache et al. 2006; Sumi et al. 2006) and higher values
(Sumi et al. 2003), although these determinations are based on
relatively small samples of microlensing events. It is important to
note that there appears to be a clear distinction between the measured
values of $\tau$ for the two commonly-used techniques:
higher values of $\tau$ are found for determinations carried out using
all stars (e.g. Alcock et al. 2000; Sumi et al. 2003), whereas lower
values are found when (as advocated by Gould [1995]) only bright stars
are used in the analysis (e.g. Popowski et al. 2005; Hamadache et
al. 2006; Sumi et al. 2006).

It was realised quite early on (e.g. Udalski et al. 1994a; Alcock et
al. 1997b) that blending is a major uncertainty in the determination of
$\tau$. Blending occurs naturally because microlensing surveys are
conducted in crowded stellar fields and, with typical ground-based
seeing, other stars can blend into the seeing disk of the lensed star
(see, for example Han 1999 and references therein). This affects the
number of potential lensed sources and also introduces uncertainties
into the fitted event parameters (e.g. Wo\'zniak \& Paczy\'nski 1997;
Han 1999). Importantly, it was proposed that the aforementioned
discrepancy between the $\tau$ measurements from all stars compared to
bright stars could be explained by blending.

Clearly the ideal way to understand this blending issue is
with high resolution $Hubble Space Telescope (HST)$ images (Han 1997),
which can be used to 
resolve any nearby blends that may be present. This technique was
adopted by Alcock et al. (2001a) in their analysis of a set of
microlensing events towards the Large Magellanic Cloud. However, in
general this method is limited due to the restrictions on the
availability of $HST$ observing time. 
Therefore, in the absence of high resolution images for each
event, the next best approach is to undertake Monte Carlo simulations
at the pixel level. Mock images can be generated based on deep $HST$
luminosity functions of the Galactic bulge. Unfortunately, since such
luminosity functions are currently only available for a very limited
number of lines of sight (e.g. Holtzman et al. 1998), one must
extrapolate their behaviour for the various bulge fields observed by
microlensing collaborations. Artificial 
microlensing stars can be injected into mock images and then convolved
into ground-based seeing. It is then possible to investigate the efficiency
of recovering microlensing events, i.e. the detection efficiency (see, for
example, Alcock et al. 2000). In order to simplify the analysis many
microlening studies concentrate on bright stars, working under the
assumption that bright stars suffer negligible blending.
However, the reliability of this assumption has recently been called
into question (see Section \ref{sec:obsevidence}).
As a result, proper consideration must be given to blending, even
when one considers microlensing of bright stars.

Monte Carlo simulations of blending have already been carried out by
various groups, mostly concentrating on the effect to the recovered
microlensing optical depth. Recent studies include Popowski et
al. (2005) and Hamadache et al. (2006), both of which argue that the
recovery of the optical depth should not be significantly biased by
the presence of blending in bright events.
The analysis presented in this paper builds upon another such work
(Sumi et al. 2005), which showed that a simulated sample of bright
microlensing events still contains many heavily blended events. We
extend the work of Sumi et al. (2005) by generalising the analysis to
fields with varying stellar density under different seeing
conditions. By doing this we aim to make broader conclusions
that go beyond any experiment-specific analysis.

The outline of the paper is as follows, in Section
\ref{sec:obsevidence} we briefly discuss some of the observational
evidence that exists to suggest that bright microlensing might not avoid
the problem of blending. The remaining sections deal with Monte Carlo
simulations that we have undertaken in order to investigate this
phenomenon: Section \ref{sec:method} describes the method; Sections
\ref{sec:general} \& \ref{sec:fit} present the results of our
simulations including the resulting distributions of event parameters;
Section \ref{sec:opdepth} investigates whether the assumption that
bright events are unblended can bias the measured value for the
optical depth; and Section \ref{sec:motion} discusses the expected
distributions of centroid motions. We conclude with Section
\ref{sec:discussion}, where we discuss the implications of our findings.

\section{Observational evidence of blending in bright events}
\label{sec:obsevidence}

Although it is often assumed that in general bright events are not
affected by blending, there is observational evidence to show that
this is not always a safe assumption. In the following section we
briefly discuss various different approaches that can be used to test
this hypothesis for observed events using ground-based data. We
provide a number of examples of bright events that exhibit significant
blending.

Throughout this section and the rest of the paper we characterise the
blending using the following parameter, $\fS$, which denotes the ratio
of the source flux to the total baseline flux, i.e.
\beq
\fS=\frac{F_{\rm source}}{F_{\rm baseline}}.
\eeq
Therefore, $\fS=1$ for the case of no blending, while
$\fS\rightarrow0$ for heavily blended events.

\subsection{General model fitting}

Numerous catalogues of gravitational microlensing events have been
published towards the Galactic bulge. Many of these contain
bright events that model fitting has suggested are blended, such as
the MACHO catalogue of Alcock et al. (2000).
This catalogue contains a subset of 37 events that are
described as `classical lensing' from which one can determine the
fitted blending parameter. Four of these events are both bright
($V<18$) and heavily blended ($\fS<0.2$): namely 95-BLG-d19,
97-BLG-d13, 97-BLG-24, 97-BLG-37. All of these four events have
reasonably well constrained values of the blending parameter. Popowski
et al. (2005) also noted that based on light curve fitting they cannot
exclude the possibility of significant blending for some events in
their clump-giant sample.

Further examples can be found in the catalogue of clump-giant EROS
events published by Afonso et al. (2003), in which two of 16 bright events
were found to display clear blending signatures (EROS-BLG-31 and 
EROS-BLG-12). In contrast, a more recent analysis of the EROS data
(Hamadache et al. 2006) finds only five of 120 clump giant events
appear to exhibit strong blending, although they acknowledge that
their paucity of blended events could be due to the limited
photometric precision of the EROS experiment.

In their analysis of the OGLE-II catalogue of bright events, Sumi et
al. (2006) found a that blending was significant for a number of their
events. According to their best-fitting models, ~38 per cent of these
bright events were actually due to lensing of much fainter
sources.

\subsection{Centroid motion}
\label{sec:obsmotion}

Another way to assess the influence of blending is to investigate the
motion of the light centroid during the event. The light centroid is
determined by the light from the lens, the lensed source, and/or
blended stars along the line of sight. Therefore, during microlensing,
the light centroid must shift towards the lensed star as it
brightens (e.g., Alard, Mao \& Guibert 1995; Goldberg 1998). If the
lens dominates the blend, the centroid shift is difficult to detect
because the lens and the lensed source are aligned to within
milliarcseconds ($\sim$ angular Einstein radius). However, if the
blending is mostly due to other blended stars, then the centroid shift
may be detectable, even in ground-based observations.

Another similar approach is to measure
the offset between the lensed source and the centroid of the light at
baseline (Han 2000); this approach can be used when dealing with
Difference Image Analysis (DIA), since in this case the location of the
lensed source can be measured to high accuracy. It is also possible to
try and locate the blend by removing the light of the lensed source
using image subtraction (Gould \& An 2002; Smith et al. 2002).

These techniques have been applied in various works. The offset
between the lensed source and the baseline centroid has been routinely
measured for various microlensing catalogues (e.g. Alcock et al. 1999;
Alcock et al. 2000; Wo\'zniak et al. 2001). In many cases there are a
significant fraction of events with offsets of 1 arcsec or greater.

\begin{figure*}
\plotone{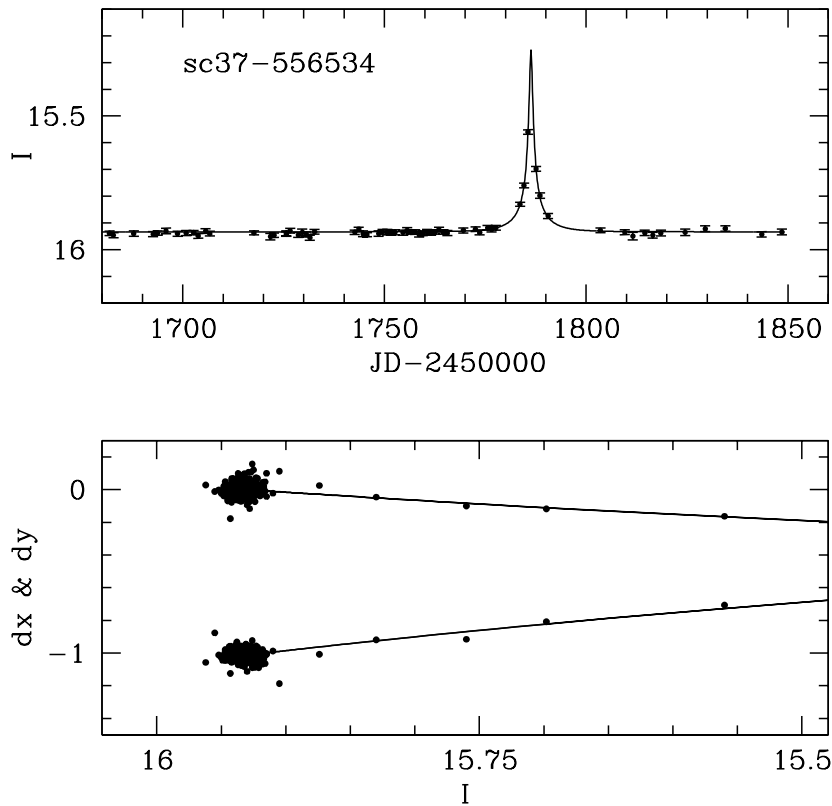}
\caption{An example of an observed case of centroid motion. The top
panel shows the light curve for this event (sc37--556534). The bottom
panel shows the corresponding centroid shift for $x$ and $y$ in pixels
(1 pixel corresponds to 0.417 arcsec) as a function of $I$-band
magnitude. The solid line indicates the best-fit model. For clarity,
the $y$ centroid position has been shifted downward by 1 pixel.}
\label{fig:centroid}
\end{figure*}

The first such detections of centroid motions have been presented in
Alard, Mao \& Guibert (1995) and Goldberg \& Wo\'zniak (1998). To
further test this effect we have examined the centroid motion for a
sample of red clump giant microlensing events from OGLE-II (Sumi et
al. 2005).
In order to constrain the offset between the lensed star and the
blend, we fit the light curve and the centroid positions
simultaneously.
Fig. \ref{fig:centroid} shows the event with the most significant
centroid shift (sc37--556534), which has a very bright baseline
magnitude ($I=15.9$ mag).
As expected, the centroid location is a strong function of the
magnitude (magnification). The offset (in pixels, where 1 pixel
corresponds to 0.417 arcsec) between the lensed star and the blend is
$\Delta x\approx0.61,\,\Delta y\approx-1.04$. This centroid shift
clearly demonstrates that blending can be important for bright
microlensing events.

\subsection{Binary lens blending}

Blending can be inferred not only for single microlensing events, but
also for binary microlensing events. In fact, the first (Udalski et
al. 1994b) and second (Alard, Mao \& Guibert 1995) binary lens
events both show significant blending. 
Detailed studies of binary lenses in OGLE-II (Jaroszy\'nski 2002)
and OGLE-III data (Jaroszy\'nski et al. 2004) 
show convincingly that blending in binary lenses is widespread; they
find that the fractions of bright events (observed $I<18$) with fitted
parameter $\fS<0.5$ are $3/5$ and $4/7$ for OGLE-II and OGLE-III,
respectively. The most 
spectacular example is sc5\_6650 from OGLE-II, for which
the lensed source is inferred to contribute only 1 per cent of
the total light even though the composite is very bright with a
baseline magnitude $I=16.18$. 

Binary light curves are very diverse and, as a result, poorly sampled
ones can often be fitted with multiple models (e.g., Mao \& Di Stefano 
1995; Albrow et al. 1999; Dominik 1999; Gaudi \& Han 2004). This
problem often renders the $\fS$ determination somewhat
uncertain. However, for binary events that undergo a caustic crossing,
a limit on the blending can often be inferred without any detailed
modelling. These binary events exhibit `U'-shaped light curves as the
lensed star enters and then 
exits from the caustics. The minimum magnification in the plateau 
must be equal to or exceed 3 (Witt \& Mao 1995). If the observed minimum
magnification within this plateau,
$A_{\rm min}$, is below 3, then the fraction of light contributed by the
source must satisfy the inequality,
\begin{equation}
\fS \le {A_{\rm min}-1 \over 2}, ~~ A_{\rm min} \le 3.
\end{equation}
For example, the OGLE-II event sc5\_6650 has an observed $A_{\rm min}
\approx 1.04$ and hence we can infer that $\fS \le 0.02$, which is fully
consistent with the blending parameter ($\fS=0.01$) obtained by
Jarosz\'nski (2002) from detailed light curve fitting. Kim (2004)
presents a more comprehensive analysis of the limits that can be
derived for binary events in the OGLE-II and OGLE-III databases; this
work shows that $5/7$ bright (observed $I<18$) caustic crossing events
have $\fS<0.5$.

It should be noted that there may be a slight bias in the blending
distribution derived from caustic crossing binary events since such
events often undergo large amplifications, which increases the
probability of observing lensing of faint source stars. However, even
given this caveat it is clear that the analysis of binary events can
provide a robust model-independent method for investigating
blending. As has been shown above, it is evident that for binary
lenses (as with single lenses) bright events can be heavily blended.

\section{Construction of a mock catalogue of blended events}
\label{sec:method}

In the remainder of this paper we undertake Monte Carlo simulations in
order to investigate the effect of blending for simulated bulge
fields with varying seeing and densities. In this section we discuss
the construction of our catalogue of mock microlensing events.

We first construct artificial images for 9 simulated fields, adopting
three different values for the density of stars and three different
values for the level of seeing. The densities of our fields are chosen
relative to the OGLE-II field sc3, which is centered on
$l=0.11^\circ$, $b=-1.93^\circ$ and has an observed surface density of
151 stars per arcmin$^2$ down to a magnitude of $I=17$. We choose
densities of 0.5, 1 and 1.5 times the density of field sc3. The three
values of seeing are 2.1, 1.05, 0.7 arcsec. Throughout this paper we
designate the field with median seeing and density (i.e. density of
field sc3 and seeing of 1.05 arcsec) as our reference field. The
details for each of our fields are summarised in Table \ref{table:blend}.
Similar to the OGLE-II experiment, each field has dimensions
14$\times$57 arcmin$^2$ and has pixel size 0.417 arcsec pixel$^{-1}$.
Field sc3 was chosen since it is a very dense stellar field close to
the Galactic centre where blending should be most significant.
Further details about the OGLE-II experiment can be found in Udalski et
al. (2000). 

It is helpful to see how the adopted characteristics for our simulated
fields compare to the important microlensing bulge surveys.
The pixel size and typical seeing values for five of the major
microlensing experiments are:
\begin{itemize}
\item OGLE-II: 0.42 arcsec pixel size and median seeing $\sim1.3$ arcsec (Sumi et al. 2006)
\item OGLE-III: 0.26 arcsec pixel size and median seeing $\sim1.3$ arcsec
\item MACHO: 0.63 arcsec pixel size and median seeing $\sim2.1$ arcsec (Popowski et al. 2005)
\item EROS-II: 0.6 arcsec pixel size and median seeing $\sim2$ arcsec (Palanque-Delabrouille et al. 1998);
\item MOA: 0.81 arcsec pixel size and median seeing $\sim2.5$ arcsec (Bond et al. 2001).
\end{itemize}
It can be seen that the range of seeing and density for our simulated fields
covers practically all bulge surveys, both current and previous.
Although our adopted pixel size (corresponding to OGLE-II) is smaller
than the non-OGLE experiments listed here, this should not be a
dominant effect. It is 
also worth noting that our choice of medium density, which
corresponds to one of the densest OGLE-II fields, is comparable to the
densest fields in both MACHO (Popowski et al. 2005) and EROS
(Hamadache et al. 2006) optical depth studies.

Our artificial fields are created following the prescription of Sumi
et al. (2006); full details of the procedure can be found in section
4.1 of their paper. We populate our field by selecting stars from the $HST$
$I$-band luminosity function of Holtzman et al. (1998). Since the OGLE
field sc3 is not coincident with the $HST$ field from Holtzman et
al. (1998), the $HST$ luminosity function must be shifted so as to
matches the observed number density of bright stars from the OGLE
sc3 field.
By combining the OGLE and $HST$ data we are also able to constrain the
bright end of the luminosity function; this region cannot be well
constrained using the $HST$ data alone as there are very few bright
stars due to saturation and the small field of view.
We account for the differential extinction across the
field using the extinction maps of Sumi (2004). Note that our
simulations are carried out solely with $I$-band data and do not
incorporate any colour information.

Using this luminosity function we populated our field with
approximately $10^7$ artificial stars down to a magnitude of $I
\approx 22$ and then convolved the image to our required level of
seeing. Our resulting mock fields are therefore fully synthetic
(i.e. artificial stars are not injected into observed fields) and have
realistic noise properties, making them effectively indistinguishable
from real images. Note that stars are placed randomly in the field and
are not placed on a regularly spaced grid.

Given this mock field we then applied the standard OGLE
photometry pipeline that is
based on DoPHOT (Schechter, Mateo, \& Saha 1993) to obtain the
`observed' magnitude of the $10^7$ artificial stars. From this we can
record the `input' magnitude of the artificial star ($\Iin$) and the
resulting `output' magnitude as measured by DoPHOT ($\Iout$). This 
`output' magnitude corresponds to the observed baseline for a
star. Ideally, we should find that $\Iout<\Iin$ for all
stars, since the `output' magnitude includes many blended
stars. However, as will be shown below, this is not true in all
cases (see Section \ref{sec:negblend}).

Given these `input' and `output' magnitudes we can calculate the 
the blending fraction, $\fS$, i.e. the fraction of the baseline flux
contributed by the source,
\beq
\fS=\frac{F_{\rm source}}{F_{\rm baseline}}=\frac{F_{\rm
    source}}{F_{\rm source} + F_{\rm blend}}=10^{(\Iout-\Iin)/2.5}.
\eeq
However, the resulting blending distribution needs to be corrected for
the detection efficiency, since heavily blended events
(i.e. $\fS\ll1$) are less likely to be detected because they require
much greater intrinsic magnifications to produce an observed increase in
magnitude.

To simulate this effect, we generated a mock catalogue of
standard microlensing light curves using the sampling and photometric
properties of the OGLE-II experiment (Udalski et al. 2000): namely,
$I$-band observations taken approximately once every few nights; bulge
season typically lasting from mid-February until the end of October;
limiting magnitude and saturation are approximately $I\approx20$ and
$I\approx11.5$, respectively. We choose to generate light curves over
a total baseline of three years and assume that each star has an equal
probability of being lensed (i.e. we assume all stars belong to the
bulge). The generation of these mock catalogues
follows the prescription given in Smith et al. (2005) although for the
purposes of this paper parallax signatures (Gould 1992) have been
neglected. Although it is well known that symmetric parallax
signatures can be confused with blending (Smith, Mao \& Paczy\'nski
2003), it would be computationally too demanding to generate and fit
parallax signatures. In any case, such symmetric parallax events
should not be common.

The event timescale is drawn from the model described in
Smith et al. (2005) and the impact parameter (in units of the Einstein
radius) is chosen uniformly between 0 and 1.5. Note that unless
otherwise mentioned, our analysis is based on events with fitted impact
parameter less than 1; we generate events with larger impact
parameters because when events are fitted with a microlensing model (see
Section \ref{sec:fit_details}) the recovered impact parameter can 
sometimes be underestimated.
Photometric errors are assumed to be Gaussian (e.g. Wo\'zniak 2000).

Once this mock catalogue has been produced we apply the selection
criteria of Sumi et al. (2005) to simulate the detection
efficiency. Essentially, these criteria test for the presence of a
constant baseline with one distinct brightening episode.

\begin{figure*}
\plotone{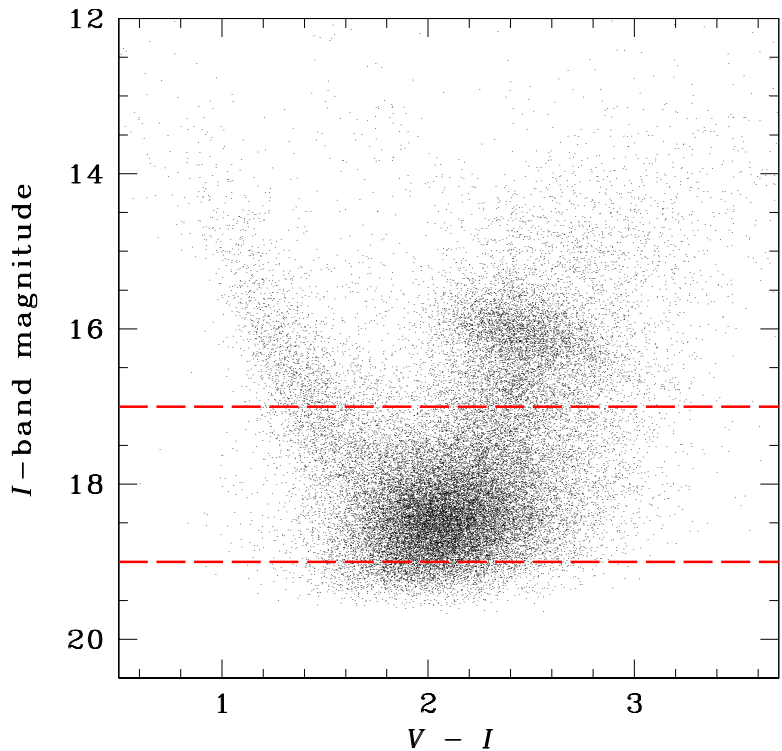}
\caption{
The observed OGLE-II colour-magnitude diagram for
field sc3. Note that this is the $observed$ field and not a
$simulated$ field, since we do not include $V$-band data in our
simulations. The horizontal lines show the cuts for all events
($\Iout<19$) and bright events ($\Iout<17$). 
The latter sample
corresponds to a brightness cut similar to that employed for red clump
giant microlensing events; for this figure the red clump region is
centred around $V-I\approx2.4$, $I\approx16$. In reality, a red clump
sample would be selected using a cut in colour to remove foreground
blue disk stars, but this is not necessary in our case since our
simulation contains no colour information and is set up so that all
stars belong to the bulge.
}
\label{fig:cmd}
\end{figure*}

We divide our mock catalogue into two samples: a full sample of all
events with baseline magnitude $\Iout<19$ and a subsample of bright
events with $\Iout<17$. This latter subsample of events
corresponds to what one would expect to see for a sample of red clump
giant microlensing events, for example. We only consider events with
$\Iout<19$ since fainter objects are too close to the limiting
magnitude of the OGLE-II experiment. These cuts are illustrated in
Fig. \ref{fig:cmd}, which shows the observed colour-magnitude diagram
for the OGLE-II sc3 field.

\section{General shape of blending distribution}
\label{sec:general}

\begin{figure*}
\plotone{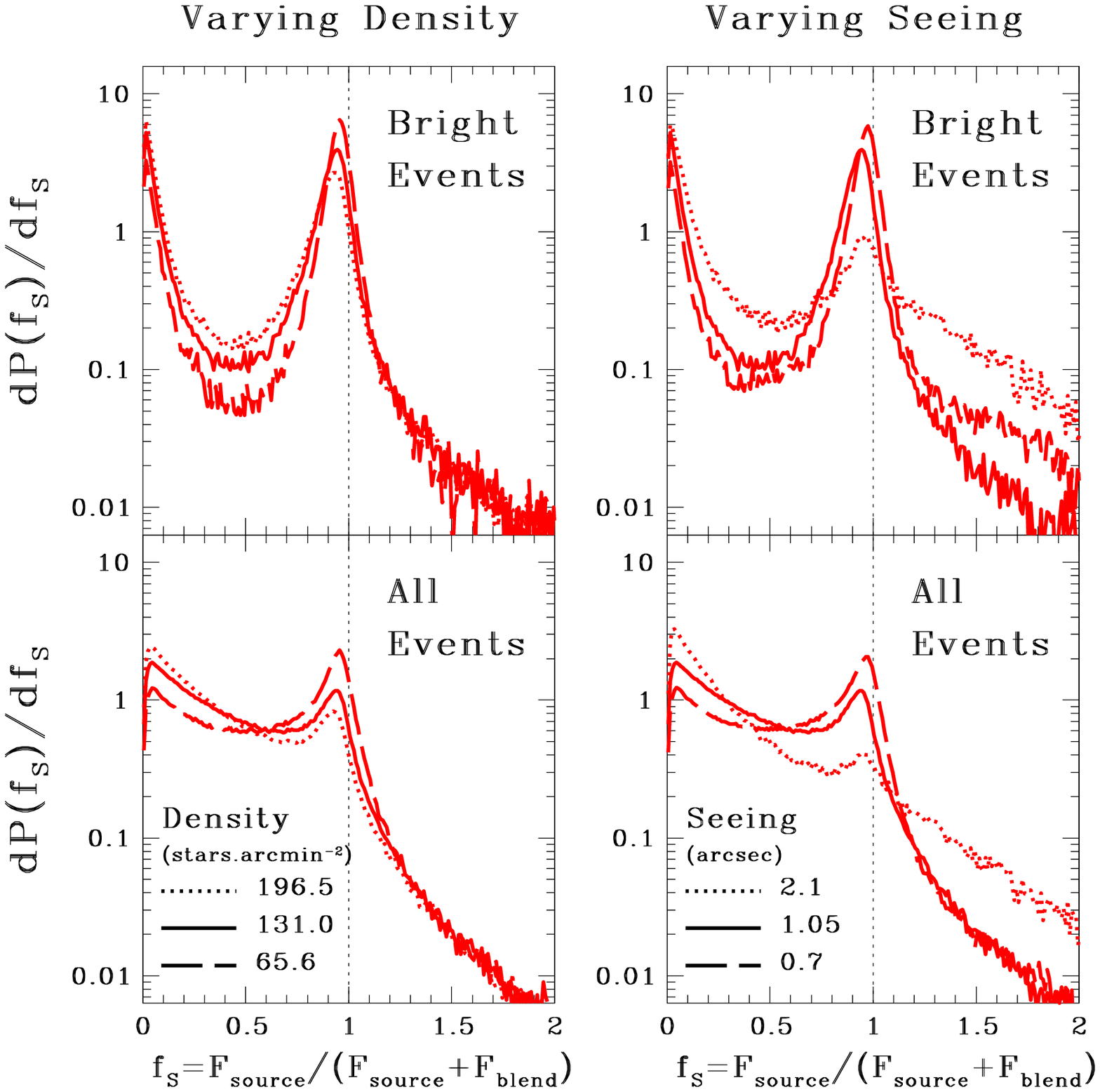} 
\caption{
The input (i.e. not fitted) blending distributions for recovered
microlensing events in our simulated
fields. The upper panel shows the distribution for a bright subset of
events ($\Iout<17$), such as red clump giants, and the lower panel shows
the distribution for all events (i.e. $\Iout<19$).
The left panel shows fields with differing stellar density
(dotted/solid/dashed correspond to high/medium/low density) while the
right panel shows fields with differing values for the seeing
(dotted/solid/dashed corresponds to low/medium/high quality seeing). 
The vertical line at $\fS=1$ denotes zero blending, which
means that all events to the right of this line exhibit negative
blending (see Section \ref{sec:negblend}).
The small-scale fluctuations in these distributions are due 
to small number statistics and are not physical.
}
\label{fig:true}
\end{figure*}

\begin{figure*}
\plotone{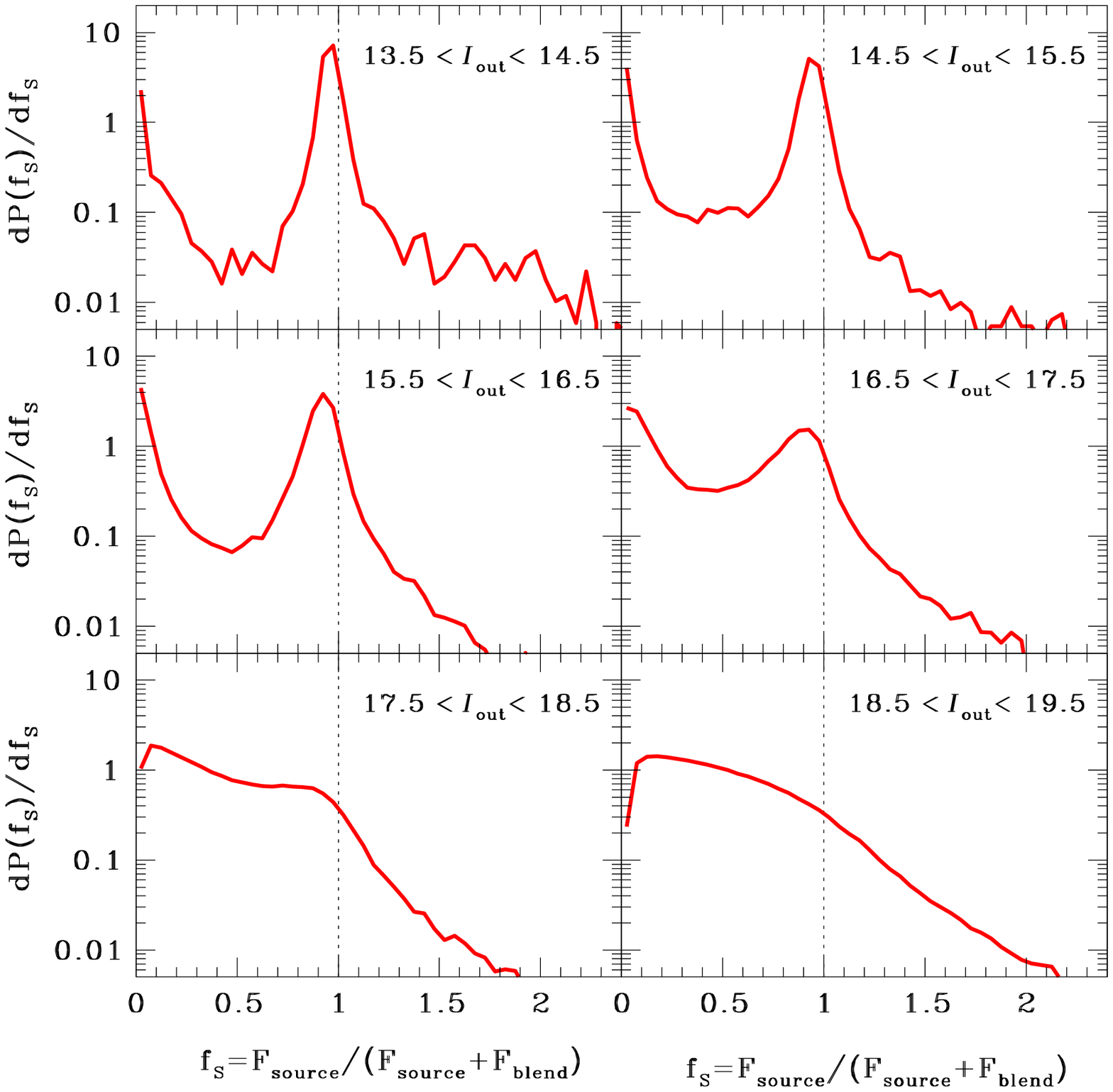}
\caption{The input (i.e. not fitted) blending distribution for
recovered microlensing events for six different output magnitudes,
$\Iout$. This figure shows the distribution for the simulated field
with medium seeing and density.
The small-scale fluctuations in these distributions are due 
to small number statistics and are not physical.
}
\label{fig:true_all}
\end{figure*}

\subsection{General description of blending distribution}

The resulting blending distributions for 5 of our simulated fields are
shown in Fig. \ref{fig:true}. In this figure we show both the full
samples of all events ($\Iout<19$) and the bright subsamples of
events with baseline magnitude brighter than $\Iout<17$ (see Section
\ref{sec:method}).

For the full sample of all events, the distributions appear relatively
flat between $\fS=0$ and $\fS=1$. The most striking feature is the
long tail of events with $\fS>1$, particularly for fields with bad
seeing. This feature corresponds to sources that are brighter than the
measured baseline (i.e. $\Iout$). These so-called `negatively
blended' events are discussed in detail in Section \ref{sec:negblend}.
As expected, there is a significant proportion of
heavily blended events; for example, for our reference field (medium
density and seeing), 57 per cent have
$\fS<0.5$ and 16 per cent have $\fS<0.1$.
For our different fields the blending distribution follows the
anticipated trend, with the better seeing and/or less dense stellar
fields exhibiting less blending (see Table \ref{table:blend}).

The shape of the blending distribution is quite different for the
bright subsample of events. There is a clear `u'-shaped distribution
with few events in the region $0.2<\fS<0.8$. It has been assumed
that bright events may be free from the problem of blending, since
any blended star would be too faint to contribute substantially to
the total flux. However, it is clear from this figure that a
significant proportion of events have $\fS<0.1$, namely 27 per
cent for our reference field. The behaviour for the other fields is very
similar, but with fewer heavily blended events for fields with better
seeing and/or lower stellar density (see Table \ref{table:blend}).

We find that in almost all cases heavily blended events occur in our
bright sample when a faint source close to a bright star is
lensed. For example, in our reference field we find that for a sample
of bright (total baseline magnitude $\Iout<17$), blended events
($\fS<0.2$), over 99 per cent have source magnitudes fainter than 17th
magnitude.
There are two competing effects that will determine how
frequently such events occur. First, it depends on the luminosity
function, i.e. how the number counts rise as the magnitude becomes
fainter. If the luminosity function is steep (as is the case for the
bright end of the luminosity function), then there are numerous faint
stars, which enhances the number of lensed cases with strong
blending. Second, in order for strongly blended events to be
observable, the intrinsic magnification must be much higher than the
nominal threshold ($A > 3/\sqrt{5}$). Given the shape of this $\fS$
distribution, it appears that the assumption that bright events are
unblended is not a sound one. We deal with the effects of this
assumption in later sections (see Section \ref{sec:fit} and
\ref{sec:opdepth}). As with the full sample of all events there is a
fraction of events with negative blending, although for this magnitude
range this fraction is smaller (see Section \ref{sec:negblend}).

In Fig. \ref{fig:true_all} and Table \ref{table:negblend_Iout} we show
how the $\fS$ distribution varies as a function of the baseline
magnitude (i.e. $\Iout$) for our reference field.
Although the overall shape of the distribution shows a significant
trend across the range of magnitudes, there is no clear trend for the
percentage of heavily blended events with $\fS<0.2$. This fact is
interesting in that it is clearly in contradiction with the argument
that bright events are free from significant blending.

\subsection{Negative blending}
\label{sec:negblend}

Positive blending is obviously induced when the lensed source is blended
with other sources that are too close to resolve under ground-based
seeing conditions. The case for negative blending is less
straightforward.
One of the first papers to discuss the issue of negative blended
fluxes was Park et al. (2004), which concerned the microlensing event
\moa. For this event they obtain a best-fit value of $\fS \approx
1.06$. Further events have been identified by various authors (e.g. Jiang
et al. 2004; Poindexter et al. 2005; Sumi et al. 2006).

Fig. \ref{fig:true} shows the extent of the issue for our simulated
fields (see also Table \ref{table:blend}). Curiously, there are no
obvious trends evident in either seeing or density. The only clear
effect is that for lowest quality seeing the problem becomes significantly 
worse, although this could be due to increased photometric noise which
is present in fields with such bad seeing.
There is also no clear trend between the samples of all events and
bright events.

We examined a number of images from our simulations that show negative
blending. In many cases we find that it arises when we have another
star close to the lensed star, and the DoPHOT photometry program
incorrectly partitioned part of the lensed star flux to the nearby
blend, yielding a negative blending for the lensed source. Another
possibility is a `hole' in the mottled distribution of faint stars
that constitutes the background flux. This arises because the DoPHOT
program assumes a constant background, while in reality the background
is contributed by unresolved faint stars which have fluctuations. If
the background close to the star is lower than the average background
(a `hole'), then the DoPHOT program will over-subtract the background,
yielding a negative blending for the lensed source. For \moa, the
best-fit deblended intrinsic source magnitude $I_0\approx14.4$ mag, in
order to produce $\fS\approx 1.06$, Park et al. (2004) required a
`hole' equivalent to the omission of a $I_0\approx17$ mag turn-off star.

To conclude, we are unable to make any definitive statements about the
nature of negative blending. We have attempted to show the empirical
effects by carrying out analyses that are as close to real surveys as
possible. It seems that the causes for negative blending are a
combination of those discussed above, namely software issues from
deblending and holes in the background, with the additional
complication of simple statistical noise.

\section{Investigation of blending-related biases in the recovery of
event parameters} 
\label{sec:fit}

\subsection{Fitting of the mock light curves and detection efficiency}

\subsubsection{Light curve fitting}
\label{sec:fit_details}

To investigate the best-fit event parameters we fit each mock light
curve with the standard 5-parameter blended microlensing model:
\beq
I(t)=I_{\rm base} - 2.5 \,{\rm log}\, [ \fS (A(t)-1) + 1],
\eeq
where the magnification, $A(t)$, is given by,
\beq
A(t) = {u(t)^2+2 \over u(t) \sqrt{u(t)^2+4}},~~
u(t) \equiv \sqrt{\u0^2 + \left(\frac{t-t_0}{\tE}\right)^2}.
\eeq
Here $\u0$ is the impact parameter in units of the Einstein radius,
$t_0$ is the time of the closest approach (i.e. maximum
magnification), and $\tE$ the event timescale. Our parameter $\tE$
corresponds to the Einstein radius crossing time; it should not 
be confused with the Einstein diameter crossing $\hat{t}=2\tE$, which
is sometimes used by the MACHO collaboration.

The best fit was found by minimizing the 
$\chi^2$ using the Minuit package from the CERN Program
Library.\footnote{http://wwwasdoc.web.cern.ch/wwwasdoc/minuit/} A
full description of the Minuit package can be found in the `Minuit
Reference Manual' (James 1994), which is also available
online.

We reject all events for which the timescales are degenerate
(i.e. those events for which errors cannot be computed by Minuit or
events with errors on $\tE$ greater than 30 per cent), since no
meaningful parameter values can be extracted from such light
curves. This mostly removes low signal-to-noise ratio (S/N) events, although a
fraction of these are heavily blended events suffering from the
well-known degeneracy between $\fS$, $u_0$ and $\tE$ (Wo\'zniak \&
Paczy\'nski 1997).

Note that it is important that we only deal with events with
well-constrained parameters since when we calculate the optical depth for
the blended fits (Section \ref{sec:opdepth}) we select our bright
samples using the fitted source magnitude, which obviously depends on $\fS$.

\subsubsection{Notation convention}
\label{sec:notation}

In the remainder of this work we adopt the following notation
conventions. We use the subscript `in' to denote an `input' property,
i.e. the true value of a parameter. For events that have been fitted with
a model incorporating blending (i.e. a 5-parameter fit) we use the
subscript `5p'. Equivalently, for a fit with a model that incorporates no
blending (i.e. a 4-parameter fit with $\fS=1$) we use the subscript
`4p'. The subscript `out' is sometimes used without an accompanying
`4p' or `5p' to denote the fitted parameter in general, i.e. when
referring to both 4- and 5-parameter fits.

For the optical depth $\tauin$ corresponds to the input
$\tau$, as given below by equation (\ref{eq:tau_in}), and $\tauout$
corresponds to the observed value obtained using the best-fitting
$\tE$ parameters for the sample of microlensing light curves (equation
\ref{eq:tau_out}, below).

We also retain the convention for magnitudes described above, namely
that $\Iin$ refers to the magnitude of a star on the `input' image
(i.e. prior to convolving to ground-based seeing) and $\Iout$ refers
to the magnitude of a star on the `output' image (i.e. the resulting
magnitude of a star from DoPHOT after the image has been convolved to
ground-based seeing). The magnitude $\IS$ denotes the magnitude of the
source star, i.e. the star that has been microlensed.

\subsubsection{Detection efficiency}
\label{sec:efficiency}

The detection efficiency is required when one wishes to compare
timescale distributions or calculate optical depths (see Section
\ref{sec:opdepth}). It is is simply the proportion of microlensing
events that are recovered as a function of the event timescale,
although it can be extended so that it becomes a function of other
parameters such as the peak magnification (e.g. Popowski et
al. 2005). We only require the efficiency for bright events with
$\Iout<17$ mag, since we will not deal with the fainter events in such
a rigorous way.

We calculate the detection efficiency separately for our blended and
unblended fits. For our unblended fits we are working under the
(not necessarily correct) assumption that events brighter than $\Iout=17$
mag are unblended. Therefore for each of our simulated fields we
generate a series of mock unblended events using the same method as
described in Section \ref{sec:method}, where the baseline magnitude
of the source is selected from our standard luminosity function.
We then apply our 4-parameter fitting routine (Section
\ref{sec:fit_details}) and calculate the fraction of events that pass
our detection criteria as a function of event timescale. In this case
the detection criteria are based on the 4-parameter fits,
e.g. $u_{\rm 0,4p}<1$.

The detection efficiency for blended events is very similar to the
unblended case, except for these we also incorporate the blended
component of the baseline flux into our event generation. This is done
using the simulations described in Section \ref{sec:method}. We then
fit our events with the 5-parameter fit and calculate the proportion
of events that pass our detection criteria. Note that the efficiency
is much reduced owing to the fact that many of these blended events
are degenerate and thus do not pass our selection criteria. It is also
important to note that for the blended fits we classify bright events
as those with source magnitude brighter than $\IS=17$ mag, which means
that events with $I_{out}<17$ can be excluded if the are sufficiently
blended.

\begin{figure*}
\plotone{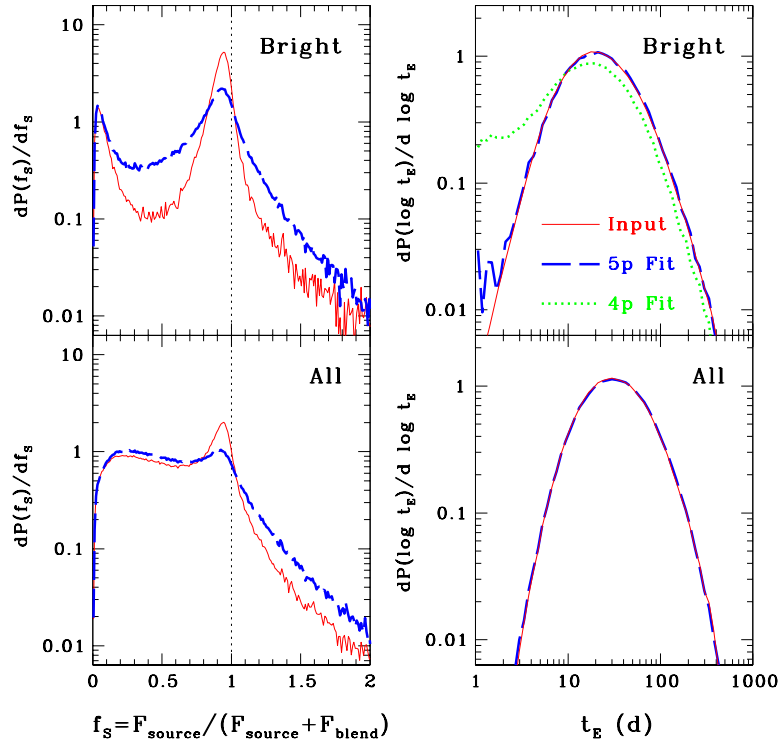}
\caption{
Input and fitted event parameter distributions for the blending
fraction (left) the event timescale (right). This figure corresponds
to the simulated field with medium stellar density and seeing; the
other fields show qualitatively similar behaviour. The solid line
represents the input event parameter distribution, the long dashed
line represents the 5-parameter fit (i.e. unconstrained blending) and the
dotted line represents the 4-parameter fit (i.e. $\fS=1$). The top
panels show the distributions for a subset of bright events
($\Iout<17$), such as red clump giants, and the bottom panels show
the distributions for all events (i.e. $\Iout<19$). The
fitted timescale distributions in the upper panel are corrected for
the detection efficiency and the solid line denotes the underlying
timescale distribution; the lower panel simply shows the distributions
of input and fitted timescales for the detected events (note that
since the two distributions in this panel are almost identical, the
two lines are coincident).
}
\label{fig:fitted}
\end{figure*}

\begin{figure*}
\plotone{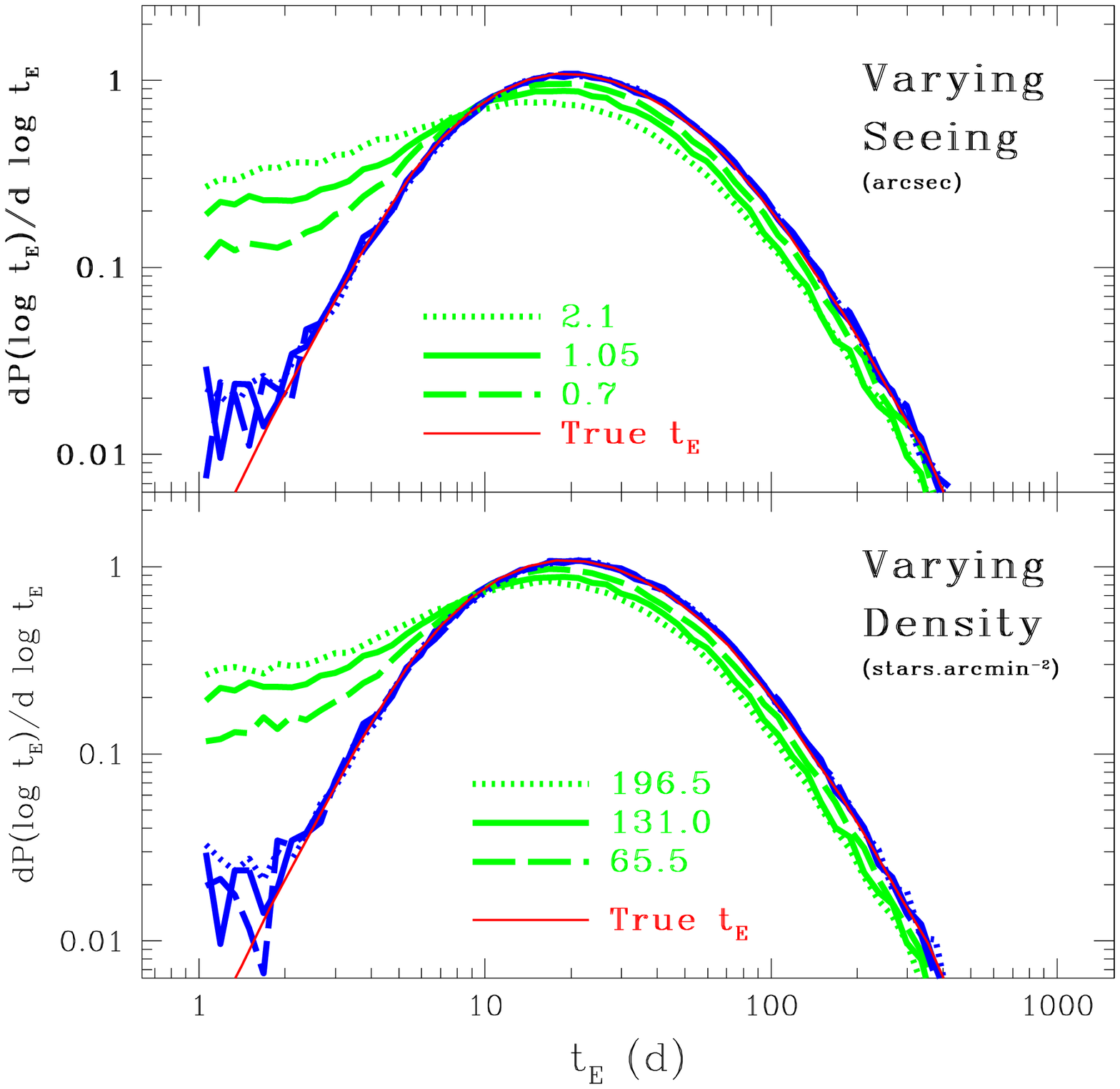}
\caption{
Input and fitted event timescale distributions for various values of
density (bottom) and seeing (top). Note that these distributions are
corrected for detection efficiency.
The thin curves denote the input timescale distribution and the thick
curves denote the fitted timescale distributions for the 5-parameter
(i.e. unconstrained blending; dark) and 4-parameter (i.e. $\fS=1$;
light) fits. In the bottom panels the dotted/solid/dashed
curves correspond to high/medium/low stellar density, while in the
top panels the dotted/solid/dashed  curves correspond to
low/medium/high quality seeing.
All curves are only for bright events ($\Iout<17$).
}
\label{fig:tEfitting}
\end{figure*}

\subsection{The effect of blending for all events}

The distributions of the fitted and input event parameters are shown in
Fig. \ref{fig:fitted} for our reference field (medium stellar
density and seeing). The bottom panels show the distributions for all
events, i.e. $\Iout<19$. For the blending parameter $\fS$, the fitted
distribution matches the input distribution reasonably well except for
a slight under-prediction of events with $\fSfivep$ just less than one
and a slight over-prediction for the number of negatively blended
events with $\fSfivep>1$. Note that this $\fSfivep$ distribution
differs from that shown in Fig. \ref{fig:true} since this does not
include degenerate events from which one cannot gain a reliable
estimate on the event parameters. This leads to a reduction in the
number of events with small $\fSfivep$ owing to the well known
degeneracy between $\fS$, $u_0$ and $\tE$ (Wo\'zniak \& Paczy\'nski
1997). In addition, such heavily blended events often have lower
signal-to-noise owing to the fact that these faint source stars
require larger amplifications to rise above the detection threshold.

The bottom-right panel shows the fitted timescale distribution for
all events ($\Iout<19$). The input timescale distribution ($\tEin$)
for the events that pass the microlensing selection criteria almost
precisely matches the distribution of fitted timescales
($\tEfivep$). The difference in $\langle \tE \rangle$ between these
two distributions is less than one per cent. Note that we do not compare
the behaviour of the $\tE$ distribution with an actual observed
distribution, such as in Alcock et al. (2000), since the nature of the
timescale distribution is subject to various uncertainties which we
do not wish to consider here. The purpose of this work is solely to
show the difference between the fitted and input $\tE$ distributions.

Although in Fig. \ref{fig:fitted} we only show the distributions for
our reference field, the other fields from our simulations exhibit
qualitatively similar behaviour.

\subsection{The effect of blending for bright events}
\label{sec:fit_bright}

The distributions of fitted event parameters for the bright sample of
events from our reference field are shown in the upper panels of
Fig. \ref{fig:fitted}. As was seen in Fig. \ref{fig:true_all}, the
distribution of blending parameter $\fSin$ shows a `u'-shaped
distribution for bright events. However, the fitted value of the $\fSfivep$
parameter shows a markedly smoother distribution, where an
under-prediction in the number of events with $\fSfivep\approx 1$
results in an over-prediction of events with $\fSfivep\approx 0.5$ and
$\fSfivep>1$.

Despite this problem, the distribution of $\tEfivep$
matches the input distribution extremely well. In the upper-right
panel we show the $\tEfivep$ distribution corrected for the microlensing
detection efficiency (see Section \ref{sec:efficiency}), along with
$\tEin$ distribution that we used to generate our events. The two
distributions are in almost perfect agreement, with the fitted 
$\langle \tE \rangle$ being recovered to within one per cent.

It is interesting to see what happens to the fitted $\tE$ distribution
when we impose the assumption that bright events are unblended. To
investigate this we fit our sample of blended mock light curves with a
model that enforces no blending (i.e. $\fS=1$), again correcting for
the detection efficiency (Section \ref{sec:efficiency}). The
upper-middle panel of Fig. \ref{fig:fitted} shows the effect of this
assumption. There is clearly a significant tail of short duration
events in the distribution of $\tEfourp$ that is not present in the
input distribution, which is due to the presence of blended events;
when blended events are fitted with an unblended model, the resulting
timescales are underestimated. However, the fit that incorporates
blending does not suffer from this problem. Because of this problem,
the $\langle \tEfourp \rangle$ is under-estimated by 23 per cent for
this field. In the next section we discuss the implications of
this assumption on the optical depth.

\begin{figure*}
\plotone{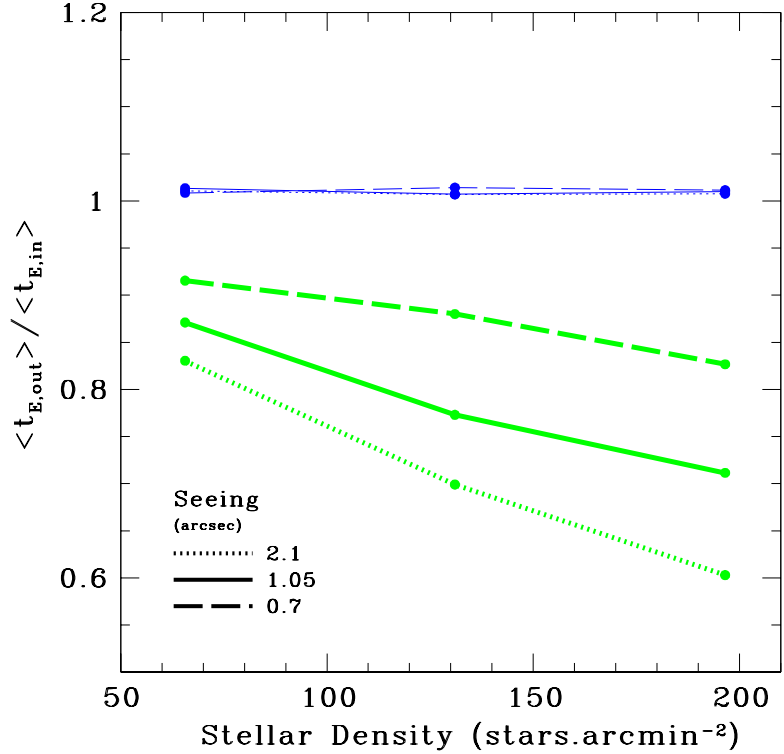}
\caption{
The ratio of (efficiency corrected) fitted vs input mean event
timescales for our simulated fields as a function of stellar
density. The thin and thick lines correspond to 5-parameter (i.e. unconstrained
blending) and 4-parameter (i.e. $\fS=1$) fits, respectively.
The dotted/solid/dashed lines denote low/medium/high quality
seeing. Note that the 5-parameter fits are very stable and therefore
the lines are practically coincident, i.e. all have $\langle \tEout
\rangle / \langle \tEin \rangle \approx 1$.
}
\label{fig:tEratio}
\end{figure*}

The qualitative behaviour for the other simulated fields is
similar to that shown for our reference field. However, the effect of the
assumption that bright events are unblended varies between fields, due
to the fact that the level of blending is dependent on the stellar
density and seeing. In Figure \ref{fig:tEfitting} we show the
$\tE$ distribution of the bright sample of events for a selection
of fields from our simulations. This figure illustrates that the
fields with a greater level of blending (i.e. those with greater
density or worse seeing) have more problems reproducing the $\tE$
distribution under the assumptions that bright events are
unblended. The corresponding dependence of $\langle \tEfourp \rangle$ with
density and seeing are tabulated in Table \ref{table:blend}
and plotted in Figure \ref{fig:tEratio}. Note that while the value of
$\langle \tEout \rangle / \langle \tEin \rangle$ is not dependent
on the seeing or density for the blended fits (for all fields
$\tE$ is recovered to within 2 per cent), there is a significant
dependence for the unblended fits.

One issue that concerned Hamadache et al. (2006) was the effect of
blending from faint stars within the seeing disc of a bright
microlensed source (as opposed to the case for which the faint
star itself is magnified).
We find that for our simulations this is a very weak effect. 
In our reference field, a sample of bright events with $\IS<17$
(i.e. source magnitude, not baseline magnitude) contains less than 0.1
per cent that are heavily blended ($\fSin<0.2$) and only around 1 per
cent with moderate blending ($\fSin<0.5$).
For these bright event with $\IS<17$, the error in the recovered
$\tEfourp$ is negligible; for our reference field the value of
$\langle\tEfourp\rangle$ is underestimated by only 2.7 per cent.
It is clear that the problems caused by blending of bright source
stars will be dwarfed by the other, much more prevalent, cause of
blending, i.e. microlensing of faint blended source stars below the
magnitude cut-off.

\section{The uncertainty in the optical depth measurement due to blending}
\label{sec:opdepth}

It is important to quantify whether the assumption that bright
microlensing events are unblended leads to any bias in the measurement
of the optical depth. As has been shown above, this assumption can
lead to a significant error in the value of the mean event
timescale. However, it has been argued that this effect cancels out
due to an overestimate in the number of events with bright
source stars (e.g. Afonso et al. 2003). It is the misidentification of the
true baseline source magnitude that causes this overestimate, since a 
sample of bright events will be contaminated by fainter source stars
that appear brighter than the magnitude cut due to blending. In this
section we test the validity of this assumption by taking a sample of
bright microlensing events from our simulations and calculating both
the input and output optical depth (i.e. both the true value and the
measured value as calculated using the conventional methods).

\subsection{Method}

\subsubsection{Input optical depth}

The first step in this procedure is to calculate the input optical
depth, $\tau_{\rm in}$, for bright source stars (i.e. $\Iin<17$).
The optical depth is the probability that a given star is magnified by
greater than $3/\sqrt{5}$. This can be estimated for a stellar field
by counting the total length of time that any of the bright stars are
magnified by greater than $3/\sqrt{5}$, divided by the total number of
bright stars and the total duration of the experiment. Since the
length of time for which any microlensed star is magnified by greater
than $3/\sqrt{5}$ is simply given by $2\tEin\sqrt{1-u_{\rm 0, in}^2}$,
the optical depth for our simulated catalogue can be calculated using
the following equation (e.g. Popowski et al. 2005),
\beq
\tau_{\rm in}=\frac{1}{T N_{\rm\star, in}} \sum_{\rm i=1}^{N_{\rm events}}
2t_{\rm E,in,i}\sqrt{1-u_{\rm 0,in,i}^2},
\label{eq:tau_in}
\eeq
where T is the duration of the microlensing experiment,
$N_{\rm \star, in}$ is the number of stars in the input field with
$\Iin<17$ (i.e. before the field has been convolved to ground based
seeing), and $N_{\rm events}$ is the total number of microlensing
events that have been generated. Note that $N_{\rm events}$ corresponds to
the number of events that have been generated, not the number of
events that have passed our detection criteria. Since we are
interested in the optical depth to bright stars, $N_{\rm events}$
corresponds to the number of events with source magnitude brighter
than this cut-off, i.e. $\Iin<17$.
Also, we only consider events that have peak
time within one of the observing seasons, i.e. $T$ does not include
the off-season. Therefore our simulations, which are based on
three year OGLE-II observations, have $T=778$ d.

\subsubsection{Output optical depth}

The input optical depth determined from equation (\ref{eq:tau_in})
must then be compared to the observed optical depth, $\tau_{\rm out}$,
which can be calculated using the conventional equation (e.g. Udalski
et al. 1994a),
\beq
\tau_{\rm out} = \frac{\pi}{2 T N_{\rm \star,out}}
\sum_{\rm i=1}^{N_{\rm events}'}
\frac{t_{\rm E,out,i}}{\epsilon(t_{\rm E,out,i})},
\label{eq:tau_out}
\eeq
where $N_{\rm \star, out}$ is the number of stars in the $output$ field
brighter than $\Iout=17$ mag, (i.e. the number of bright stars detected by
DoPHOT once the field has been convolved to ground based seeing) and
$N_{\rm events}'$ is the total number of microlensing events that have
passed the event detection criteria and have well-constrained values
for the fitted timescale (see Sections \ref{sec:method} \&
\ref{sec:fit_details}). $N_{\rm events}'$ only incorporates events with
baseline magnitude $\Iout<17$ and impact parameter
$u_{\rm 0,out} < 1.0$. The detection efficiency corresponds to the
fraction of events that pass our event detection criteria (see Section
\ref{sec:efficiency}). Note that equation (\ref{eq:tau_out}) is not
the only way one can estimate the optical depth; for example, Popowski
et al. (2005) advocate a slightly different approach that incorporates
the maximum amplification into the detection efficiency.

We estimate two different values of $\tau$: the optical depth
estimated under the assumption that bright events are unblended
($\tau_{\rm out, 4p}$) and the value estimated when blending is incorporated
into the fit ($\tau_{\rm out, 5p}$). Each of these $\tau$ estimates
are based on samples of events created using the respective detection
critera, e.g. for $\tau_{\rm out, 4p}$ the detection criteria requires
$u_{\rm 0,4p}<1$, while for $\tau_{\rm out, 5p}$ the detection
criteria requires $u_{\rm 0,5p}<1$. Note that for our blended fits, we
only retain events for which the source magnitude is brighter than
$I_{\rm S, 5p}=17$ mag, not the total baseline magnitude.

\subsection{Results}

\begin{figure*}
\plotone{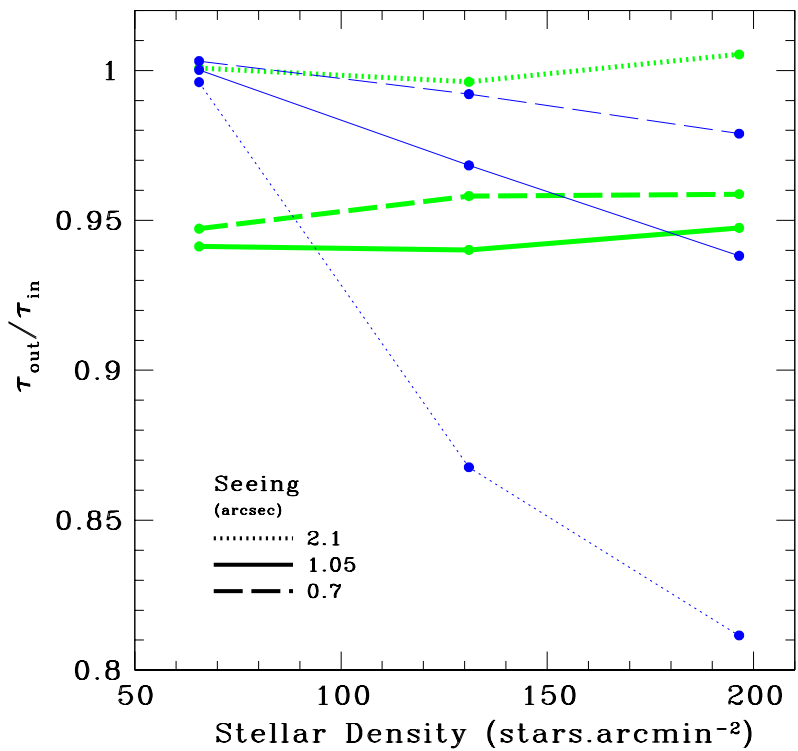}
\caption{
The ratio between the input and output optical depth for 4-parameter
(i.e. $\fS=1$; thick) and 5-parameter (i.e. unconstrained blending;
thin) fits for our simulated fields. The dotted/solid/dashed lines
denote low/medium/high quality seeing.
}
\label{fig:opdepth}
\end{figure*}

To assess the reliability of recovering $\tau$ for our two methods, we
calculate the ratio $\tau_{\rm out}/\tau_{\rm in}$ for each of
our simulated fields. This information is given in Table
\ref{table:blend} and is shown in Fig. \ref{fig:opdepth}.

In general, the 5-parameter blended fits produce good agreement between
$\tau_{\rm out}$ and $\tau_{\rm in}$, with an error of less
than 20 per cent. Note that the error is less than 10 per cent for all
fields except those with lowest quality seeing (i.e. 2.1 arcsec).

This behaviour can be understood as follows.
The dominant effect for the 5-parameter fits is the error in the
estimation of $N_{\rm \star, out}$. This value is 
determined from the number of observed stars with $\Iout<17$ mag, but
since a number of stars on the image are blended we find that that
$N_{\rm \star, out} >  N_{\rm \star, in}$. Blending can affect the
recovery of $N_{\rm star}$ in two ways: firstly, faint stars can be
brought into the bright regime by being blended with other faint
stars; and secondly, a pair of bright objects can be unresolved and
hence be detected as only one object. These are clearly two competing
effects, yet the fact that $N_{\rm \star, out} >  N_{\rm \star, in}$
shows that the former effect is dominant, reflecting the steepness
of the luminosity function at the main sequence turn-off.
In general we find that $N_{\rm \star, out}$ overestimates the number
of stars by around 10 per cent. However, for the fields with worst
seeing the error in $N_{\rm \star, out}$ can reach as high as 36 per
cent, which explains why the error in $\tau_{\rm out}$ increases for
fields with worse seeing.
In theory one could correct for this issue using simulations, although
in practise this may be difficult to estimate accurately due to
certain observational effects (Sumi et al. 2006). 

However, even if we could correct for this issue the problem is not
fully resolved. For example, if we use $N_{\rm \star, in}$ in equation
(\ref{eq:tau_out}) the 5-parameter fit still overestimates the
optical depth by around 10 per cent for all fields. This is in part
due to minor effects that influence the recovery of
$\tau_{\rm out}$, such as incorrectly estimated source magnitude and
impact parameter from the model fits. In addition to this, there is an
error associated with the fact that to obtain equation
(\ref{eq:tau_out}) from equation (\ref{eq:tau_in}), one assumes that
the impact parameter is uniformly distributed. In practise this is not
entirely true since events with impact parameter close to 1 have lower
detection efficiency than events with larger magnification
(i.e. smaller $\u0$). This problem can be overcome by using an
efficiency that is a function of both $\tE$ and $\u0$ (e.g. Popowski
et al. 2005) or by introducing a correction factor determined from
Monte Carlo simulations (such as those presented in this work).

The case for the 4-parameter fits is somewhat less subtle.
As has been stated above in Section \ref{sec:fit_bright}, when a
sample of bright events are erroneously assumed to exhibit no
blending, the fitted timescales are shifted towards smaller values
(see Fig. \ref{fig:fitted} and Table \ref{table:blend}). However, the
effect on the optical depth is counterbalanced by the increase in the
number of observed events, i.e. a number of events with source star
magnitudes fainter than the cut-off are included in the sample due to
blending (see, for example, Alcock et al. 1997b). 
For example, in our reference field we find that 22 per cent of bright events
are actually caused by fainter sources with $I_{\rm S, in}>17$. In
our extreme case of lowest quality seeing and highest density
this fraction rises to 47 per cent.
Owing to the cancellation of these two effects, from Table
\ref{table:blend} we see that the error in the unblended optical
depth is less than 7 per cent for all of our simulated fields, even
for those with highest stellar density or worst quality seeing. In most
cases the recovered optical depth is slightly lower than the input
value, but such an error can be considered negligible compared to the
statistical errors that will be present in any current real-life
experiment.
It should also be noted that there are a number of additional factors
that affect the final optical depth, such as the fact that many 
events with $u_{\rm 0,in}<1$ have $u_{\rm 0,4p}>1$ after fitting with
a model assuming no blending (see Fig. \ref{fig:fitted}). As was
discussed in Section \ref{sec:fit_bright}, there is the issue of
bright source stars being blended with nearby faint stars,
but we have found this to have only limited influence.
The problem described above of incorrectly determining $N_{\rm \star,
  out}$ also affects the 4-parameter fits, although in this case it is
not the dominant factor. However, this will still bias the optical depth
towards smaller values. 

\section{Centroid motion and centroid-source offset for blended
  events}
\label{sec:motion}

As has been shown in Section \ref{sec:obsmotion}, it is possible to
identify blended microlensing events through their centroid motion, or
equivalently the offset between the baseline centroid and the lensed
source.

\begin{figure*}
\plotone{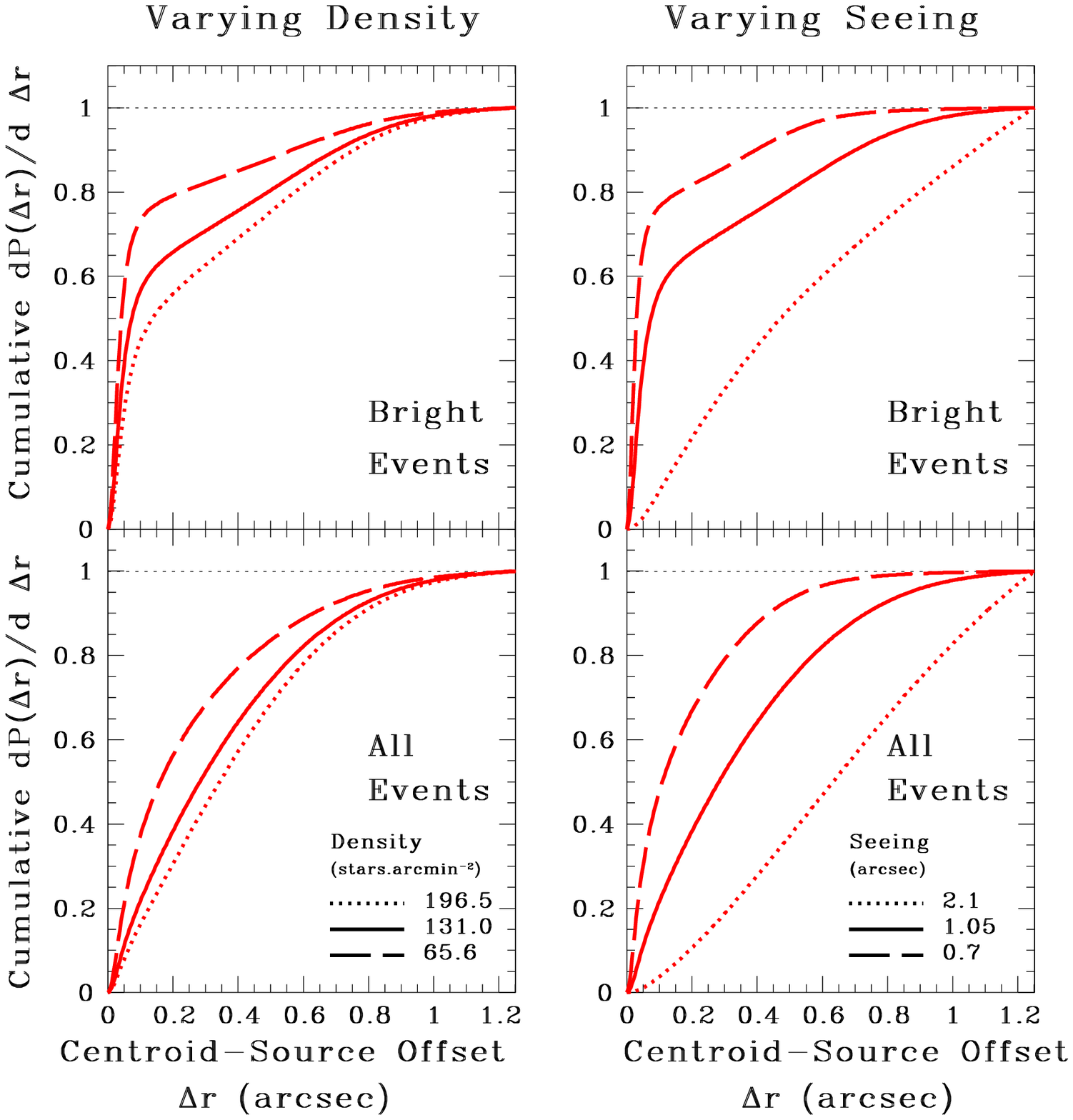}
\caption{The distributions for the offset between the baseline
centroid and the lensed source for microlensing events in our
simulated fields. The upper panels show
the distribution for a bright subset of events ($\Iout<17$),
such as red clump giants, and the lower panels show the distribution
for all events ($\Iout<19$).
The left panels show fields with differing stellar density
(dotted/solid/dashed correspond to high/medium/low density) while the
right panel shows fields with differing values for the seeing
(dotted/solid/dashed corresponds to low/medium/high quality seeing).  
Note that in our simulations one pixel corresponds to 0.417 arcsec.
}
\label{fig:offset}
\end{figure*}

\begin{figure*}
\plotone{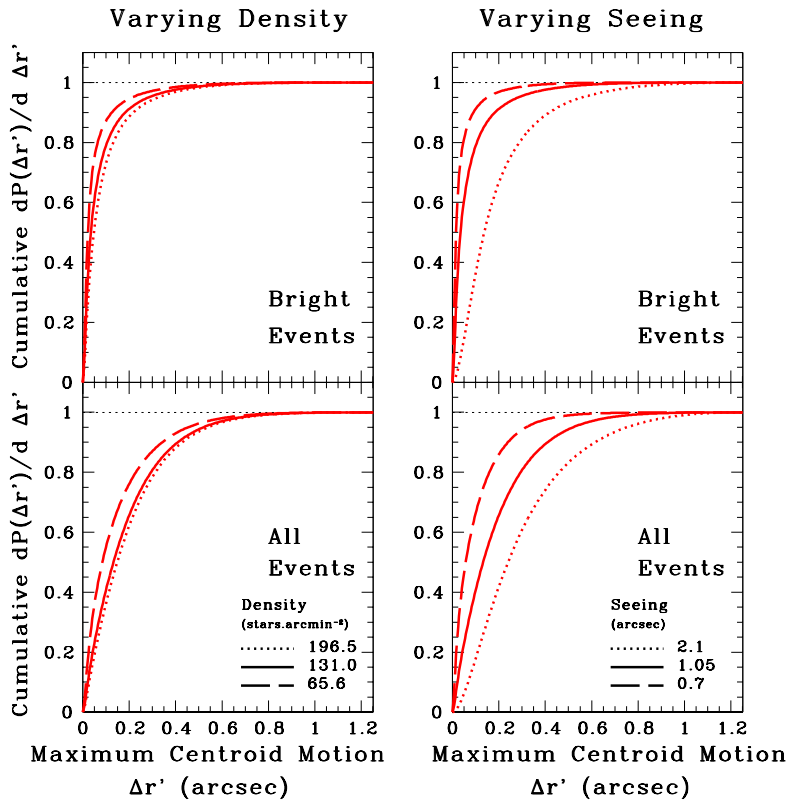}
\caption{As Fig. \ref{fig:offset} except this shows the
centroid motion for microlensing events in our simulated fields. The
centroid motion corresponds to the motion that would be observed if
the event were observed at the peak magnification (see Section
\ref{sec:motion}).
}
\label{fig:cmotion}
\end{figure*}

In Fig. \ref{fig:offset} we show the distribution of the offset
between the baseline centroid and the lensed source for our simulated
fields,
\beq
\label{eq:offset}
\Delta r = \left|
\frac{({\mathbf r}_{\rm source}-{\mathbf r}_{\rm blend})F_{\rm blend}}
{F_{\rm blend} + F_{\rm source}}\right|,
\eeq
where ${\mathbf r}_{\rm blend}$ and ${\mathbf r}_{\rm source}$ denote
the location of the blend and source centroid, respectively, and
$F_{\rm blend}$ and $F_{\rm source}$ denote the blend and source flux,
respectively.
From Fig. \ref{fig:offset} it is clear that for almost all fields a
significant fraction of events exhibit noticeable offsets of more than
0.2 arcsec. Although $\Delta r$ is reduced for the
bright samples of stars, there is still greater than 10 per cent
with $\Delta r > 0.2$ arcsec. The amount of offset is
particularly severe for the fields with low quality seeing; for
example, for field s1d2, which has a seeing of 2.1 arcsec, more than
80 per cent of events have $\Delta r > 0.2$ arcsec, even for
the bright sample.

However, one must interpret this figure with caution because one cannot
simply deduce that all of the events with large offset are heavily
blended. For fields with low quality seeing (2.1 arcsec), as many as
50 per cent of bright, $unblended$ ($0.7<\fS<1.3$) events have $\Delta
r > 0.2$ arcsec. Fortunately this problem is reduced for
fields with better quality seeing, such as our reference field (seeing 1.05
arcsec); of the bright events in this field with $\Delta r > 0.2$
arcsec, only 8 per cent show no significant blending 
(i.e. $\fS>0.5$). However, the problem remains for low quality seeing,
even in the sample of bright stars. Therefore, it is apparent that if
one wishes to use such information to quantify the level of blending,
then this must be used with caution. In particular, it is unlikely
that any meaningful results can be obtained for cases of low quality
seeing (e.g. 2.1 arcsec).

It is also possible to calculate the maximum centroid motion ($\Delta
r'$) that would be observed for the events in our simulations. Given
that the location of the centroid can be expressed as,
\beq
{\mathbf r}(t)=\frac{{\mathbf r}_{\rm blend}F_{\rm blend}+{\mathbf r}_{\rm source}F_{\rm source}A(t)}
{F_{\rm blend} + F_{\rm source} A(t)},
\eeq
where $A(t)$ denotes the magnification at time $t$. Given
this formula it is trivial to derive the expression for the maximum
centroid motion,
\beq
\label{eq:cmotion}
\Delta r' = 
\left| \frac
{F_{\rm blend}F_{\rm source}(A_{\rm peak}-1)
({\mathbf r}_{\rm source}-{\mathbf r}_{\rm blend})}
{(F_{\rm blend}+F_{\rm source}A_{\rm peak})
(F_{\rm blend}+F_{\rm source})}\right|,
\eeq
where $A_{\rm peak}$ is the maximum magnification. This equation can
equivalently be expressed in terms of the centroid-sourse offset
($\Delta r$, given in equation \ref{eq:offset}),
\beq
\Delta r'=\Delta r \frac
{F_{\rm source}(A_{\rm peak}-1)}
{F_{\rm blend} + A_{\rm peak} F_{\rm source}},
\eeq
Clearly this maximum centroid motion corresponds to the difference
between the location of the centroid at baseline and at peak
magnification, and hence is an upper limit to the observed motion
since the event may not be observed exactly at peak magnification.

The cumulative distribution for $\Delta r'$ is shown
in Fig. \ref{fig:cmotion}. As expected, this figure shows similar
behaviour to the corresponding plot of baseline centroid-source
offsets, although the offsets are somewhat
smaller. For example, for our reference field (medium seeing and
density) 30 per cent of all events and less than 10 per cent of bright
events have $\Delta r'> 0.2$ arcsec.

\section{Discussion}
\label{sec:discussion}

This paper has provided a detailed investigation into the nature of
blending in gravitational microlensing experiments towards the Galactic
bulge. As we have discussed in Section \ref{sec:obsevidence}, it is
clear that for some bright events blending can be significant. Our
simulations also support this claim. In Section \ref{sec:general} we
showed the true underlying blending distributions for our simulated fields,
all of which exhibit blending to varying degrees. We can conclude that 
the dominant source of blending is from lensing of faint source stars,
rather than lensing of bright source stars blended with nearby fainter
stars.

However, the results from our simulations (Sections \ref{sec:method}
-- \ref{sec:opdepth}) indicate that this issue may not be as
troublesome as one might fear. Although the event timescale is
unquestionably affected, the optical depth determinations in most
cases are robust across a range of seeing and density. The fact that
the optical depth is reliably recovered even from a sample of blended
events is particularly important; however, this reliable recovery
appears to rely on a coincidental cancelling out of two factors (an
underestimation in the timescale and a corresponding overestimation
of the number of events). This finding is not new, but to discover
that it holds for various values of seeing and density in our
simulation is reassuring. Clearly we do not advocate that our results
should be applied directly to any real experiments, since the details
will depend on the individual experimental set-up and event detection
criteria.

The results in Section \ref{sec:fit} are of slightly greater concern in
relation to the recovery of the event timescale. If we (erroneously)
assume that our bright events are unblended then this clearly results
in an underprediction of the event timescale of between 10 and 40 per
cent. It has been argued that the extent of this problem can be
reduced if one only deals with high amplification events. For example,
the recent papers of Popowski et al. (2005) and Hamadache et
al. (2006) advocate only using events with amplification greater than
1.5 and 1.6, respectively. If we apply this restriction to our
simulations then the level of discrepancy between the fitted and input
timescale is reduced slightly; for example, if we apply the
restriction that the $A>1.6$ for our reference field we find that the
ratio between the fitted and input $\tE$ becomes 0.81, which is only a
slight improvement compared to the ratio 0.77 that was found for
events with $A>1$. Although this does improve the situation, it is far
from resolving the problem. This underprediction could help to explain
the discrepancy in $\langle\tE\rangle$ between Popowski et al. (2005;
$\langle\tE\rangle=15\pm15$ d) and Sumi et al. (2006;
$\langle\tE\rangle=28.1\pm4.3$ d), since the former work assumes that
their bright events are unblended while the latter work incorporates
blending into their model fitting. However, there are large errors on
$\langle\tE\rangle$ for Popowski et al. (2005) and hence the
discrepancy is only very weak. It is also interesting to note that
another recent clump giant survey that assumes no blending (Hamadache
et al. 2006; $\langle\tE\rangle=28.3\pm2.8$ d) finds a value that is
in good agreement with Sumi et al. (2006), although owing to their
larger spatial coverage one might expect them to find a larger
$\langle\tE\rangle$ (e.g., Wood \& Mao 2005).

An aspect of blending that has been ignored in this work is that of
colour. Clearly if the colour of the blend and the source are
different, then the microlensing amplification will be chromatic
(e.g., Kamionkowski 1995; Buchalter, Kamionkowski, \& Rich 1996). The
strategy for the OGLE experiment has been to take practically all
observations in the $I$-band, with only very limited sampling in the $V$-band
for determining baseline colour information. However, EROS and MACHO
both have colour information, although in their optical depth
papers (Popowski et al. 2005, Hamadache et al. 2006) colour
information was not used to discriminate individual blended events
(although the former work did note that the colour variation of their
sample was consistent with expectations for a sample of unblended events).

Although the work presented here has dealt with the issues relating to
blending from stars that are coincidentally located close to a
microlensed source, there is an additional factor that we have not
considered, namely the luminosity of the lens. By definition every
microlensed source must be coincident with a lens star. Various
works have attempted to address how this fact affects the amount of
blending and the recovery of parameters from a theoretical
standpoint (e.g., Kamionkowski 1995; Nemiroff 1997; Han 1998). In
addition, for one event towards the LMC (Alcock et al. 2001b) and at
least one event towards the Galactic bulge (Koz{\l}owski et al., in
preparation) the lens and source have actually been resolved, implying
that for these events the lens contributes a finite amount of flux to
the total blended light. If more events can be found for which the
lens and source are resolvable, it will provide a promising way to
empirically quantify the effect of lens blending.
Since the Monte Carlo analysis presented here (Sections \ref{sec:method}
-- \ref{sec:opdepth}) has not incorporated this lens flux, our work
can be considered to underestimate the effect of blending.

One obvious deficiency in this work (and all other similar work) is
the lack of available deep luminosity functions for the different
lines of sight towards the bulge. Deep HST luminosity functions, such
as the Holtzman et al. (1998) one, would be very valuable if they were
available across different bulge fields. A project to tackle this
issue is currently underway (Koz{\l}owski et al., in preparation).

Data from projects such as OGLE-II can (and have) been used for
additional purposes not directly related to microlensing, such as
producing extinction maps (e.g. Sumi 2004) and constraining the
parameters of the Galactic bar through red clump giant number counts
(e.g. Rattenbury et al. 2007). These studies will also be affected by
the problems of stellar crowding. For example, any work that wishes to
determine the centre of the red clump giant region will find it
systematically shifted towards brighter magnitudes due to
blending. From our simulations we can estimate this effect; by fitting
our observed luminosity function with a power-law plus Gaussian (where
the Gaussian represents the red clump stars; see, for example,
equation 4 of Sumi 2004) one can easily determine the shift. For our
reference field we find that the Gaussian is centred on an I-band
magnitude of $16.062 \pm 0.004$ and $15.918 \pm 0.005$ for the input
(i.e. HST) and output (i.e. after convolving to ground-based seeing)
luminosity functions, respectively, i.e. a shift of 0.144 mag. Such a
shift can have an important effect in applications such as those
mentioned above. However, for the issue of extinction, one is often
more interested in the colour of the centre of the red clump
region. Since we do not have any colour information in our simulations
we cannot address this point, but from Fig. \ref{fig:cmd} it can be
seen that if the blend is caused by a star that is less than $\sim 2$
magnitudes fainter than the red clump region, then there will be a
systematic blueward shift.

In conclusion, we confirm that blending has only a limited impact on the
recovery of optical depth for the current generation of microlensing
experiments.
In Section \ref{sec:opdepth} we have shown that the recovered optical
depth is probably reliable to within $\sim 10$ per cent.
However, it is now becoming clear that with the improved accuracy of
future experiments, microlensing surveys will not be able to hide from
the issue of blending. Such work must undertake a detailed and
thorough treatment of this effect otherwise their results could be
subject to significant bias.

\section*{Acknowledgments}

The authors are indebted to the referee, Andy Gould, for providing
useful suggestions and comments.
We also wish to thank Dave Bennett for raising the issue of blending
related bias in the estimation of extinction.
MCS acknowledges financial support by PPARC and the Netherlands
Organisation for Scientific Research (NWO).
PW was supported by the Oppenheimer Fellowship at LANL.
TS acknowledges financial support from the JSPS and also benefited
from the financial help of the visitor's grant at Jodrell Bank.
This work was partially supported by the European Community's Sixth
Framework Marie Curie Research Training Network Programme, Contract
No. MRTN-CT-2004-505183 `ANGLES'.

{}

\begin{landscape}
\begin{table}
\begin{tabular}{cccccccccccccccc}
\hline
Seeing & Input & ``Observed''& Stars per & Magnitude &&
$\fS<0.2$ & $\fS<0.5$ & $0.8<\fS<1.2$ & $\fS>1.5$ &&
$\langle \tEfourp \rangle / \langle \tEin \rangle$ &
$\langle \tEfivep \rangle / \langle \tEin \rangle$ &&
$\tau_{\rm out, 4p}/\tau_{\rm in}$ &
$\tau_{\rm out, 5p}/\tau_{\rm in}$ \\
& stellar density & stellar density & ``seeing disc'' & cut &&&&&&&&&&& \\
(arcsec) & (stars.arcmin$^{-2}$) & (stars.arcmin$^{-2}$) & (stars) & && (\%) & (\%) & (\%) & (\%) &&&&&& \\
\hline
2.1 & 66.7 & 74.1 & 231.0 & $\Iout<19$ &&
30.5 &  58.0 &  18.7 &   3.7 &&
- & - && - & -\\
&&&& $\Iout<17$ &&
32.1 &  38.2 &  41.5 &   7.1 &&
0.830 & 1.011 && 1.001 & 0.996\\

2.1 & 133.1 & 167.6 & 461.0 & $\Iout<19$ &&
45.5 &  70.8 &  11.4 &   3.6 &&
- & - && - & -\\
&&&& $\Iout<17$ &&
47.6 &  57.5 &  21.7 &   6.6 &&
0.699 & 1.007 && 0.996 & 0.868\\

2.1 & 198.6 & 263.5 & 687.9 & $\Iout<19$ &&
54.8 &  76.4 &   9.3 &   2.8 &&
- & - && - & -\\
&&&& $\Iout<17$ &&
56.3 &  68.0 &  15.2 &   4.3 &&
0.603 & 1.008 && 1.005 & 0.812\\\\

1.05 & 66.7 & 72.2 & 57.8 & $\Iout<19$ &&
19.3 &  38.9 &  38.9 &   0.8 &&
- & - && - & -\\
&&&& $\Iout<17$ &&
19.4 &  21.8 &  72.8 &   0.8 &&
0.871 & 1.014 && 0.941 & 1.000\\

1.05 & 133.1 & 149.0 & 115.3 & $\Iout<19$ &&
29.9 &  56.6 &  22.8 &   0.8 &&
- & - && - & -\\
&&&& $\Iout<17$ &&
31.7 &  35.9 &  55.1 &   0.8 &&
0.773 & 1.007 && 0.940 & 0.968\\

1.05 & 198.6 & 229.2 & 172.0 & $\Iout<19$ &&
37.4 &  65.5 &  16.7 &   0.7 &&
- & - && - & -\\
&&&& $\Iout<17$ &&
40.2 &  46.1 &  42.3 &   0.8 &&
0.711 & 1.010 && 0.947 & 0.938\\\\

0.7 & 66.7 & 72.3 & 25.7 & $\Iout<19$ &&
13.2 &  27.0 &  53.4 &   0.9 &&
- & - && - & -\\
&&&& $\Iout<17$ &&
12.1 &  15.2 &  74.7 &   2.4 &&
0.915 & 1.009 && 0.947 & 1.003\\

0.7 & 133.1 & 146.2 & 51.2 & $\Iout<19$ &&
19.3 &  39.4 &  38.0 &   0.8 &&
- & - && - & -\\
&&&& $\Iout<17$ &&
19.2 &  22.0 &  69.9 &   2.1 &&
0.880 & 1.014 && 0.958 & 0.992\\

0.7 & 198.6 & 220.4 & 76.4 & $\Iout<19$ &&
25.1 &  50.0 &  27.8 &   0.7 &&
- & - && - & -\\
&&&& $\Iout<17$ &&
25.8 &  29.3 &  61.7 &   1.7 &&
0.827 & 1.011 && 0.959 & 0.979\\

\hline

\end{tabular}
\caption{
Quantitative comparison of all nine simulated fields. The first four
columns describe the setup of our simulated fields, while the
remaining columns show the fraction of blended events (columns 6--9),
the ratio between the input and fitted $\tE$ (columns 10 \& 11) and the
ratio between the input and fitted $\tau$ (columns 12 \& 13).
The subscript 4p indicates a four-parameter fit (i.e. $\fS=1$), while
the subscript 5p indicates a five-parameter fit with unconstrained blending
(see Section \ref{sec:notation}).
The fractions of blended events correspond to the input blending
distributions, not the fitted distributions. Note that $\fS>1$
corresponds to negatively blended events. Stellar densities are quoted
for stars with $I<17$.}
\label{table:blend}
\end{table}
\end{landscape}

\begin{table}
\begin{tabular}{lccccc}
\hline
&& $\fS<0.2$ & $\fS<0.5$ & $\fS>1.0$ & $\fS>1.5$ \\
&& (\%) & (\%) & (\%) & (\%) \\
\hline
$13.5<\Iout<14.5$ && 14.4 & 15.8 & 15.6 & 2.3\\
$14.5<\Iout<15.5$ && 25.2 & 28.1 & 9.3 & 0.8\\
$15.5<\Iout<16.5$ && 33.0 & 36.0 & 8.4 & 0.5\\
$16.5<\Iout<17.5$ && 37.6 & 49.5 & 7.6 & 0.9\\
$17.5<\Iout<18.5$ && 31.3 & 62.8 & 5.5 & 0.6\\
$18.5<\Iout<19.5$ && 21.3 & 58.6 & 8.1 & 1.3\\
\hline
\end{tabular}
\caption{Percentage of blended events as a function of
$\Iout$ for our field with medium quality seeing and medium
density. These percentages correspond to the input blending
distributions ($\fSin$), not the fitted distributions.}
\label{table:negblend_Iout}
\end{table}

\end{document}